\documentclass[twocolumn]{svjour3}
\usepackage[english]{babel}
\usepackage[utf8]{inputenc}
\usepackage[T1]{fontenc}
\usepackage{amsmath}
\usepackage{mathtools,amssymb}
\usepackage{graphicx}
\usepackage{bm}
\usepackage{gensymb}
\usepackage{color}
\usepackage{hyperref}
\usepackage{upgreek}
\usepackage{scalerel}
\usepackage{pgffor}
\usepackage{pdfpages}
\usepackage{float}
\usepackage{appendix}
\usepackage{physics} 
\usepackage{lscape}   
\usepackage{silence}
\usepackage{soul,xcolor} 
\usepackage[normalem]{ulem}
\setstcolor{red} 
\journalname{Journal of Superconductivity and Novel Magnetism}
\usepackage[style=phys,articletitle=false,biblabel=brackets,chaptertitle=false,pageranges=false]{biblatex}
\bibliography{biblio}

\usepackage{comment}  
\makeatletter
\AtBeginDocument{\let\LS@rot\@undefined}
\makeatother

\newcommand{\editE}[1] {\textcolor{black}{#1}}

\begin{document}

\title{Anomalous proximity effect under Andreev and Majorana bound states}

\author{Eslam Ahmed         \and
        Yukio Tanaka        \and
        Jorge Cayao
}


\institute{Eslam Ahmed \at
              Department of Applied Physics, Nagoya University, Nagoya 464--8603, Japan
              \email{jy.63s.7433@s.thers.ac.jp }           
           \and
           Yukio Tanaka \at
              Department of Applied Physics, Nagoya University, Nagoya 464--8603, Japan
              \email{o47264a@nucc.cc.nagoya-u.ac.jp}
            \and
            Yukio Tanaka \at
              Research Center for Crystalline Materials Engineering, Nagoya University, 464-8603 Nagoya, Japan
              \email{o47264a@nucc.cc.nagoya-u.ac.jp}
            \and
            Jorge Cayao \at
              Department of Physics and Astronomy, Uppsala University, Box 516, S-751 20 Uppsala, Sweden
              \email{jorge.cayao@physics.uu.se}
}

\date{Received: date / Accepted: date}
\maketitle
\begin{abstract}
We theoretically study the anomalous proximity effect in a clean normal metal/disordered normal metal/superconductor junction based on a Rashba semiconductor nanowire model. The system hosts two distinct phases: a trivial helical phase with zero-energy Andreev Bound States and a topological phase with Majorana Bound States. We analyze the local density of states and induced pair correlations at the edge of the normal metal region. We investigate their behavior under scalar onsite disorder and changing the superconductor and disordered region lengths in the trivial helical and topological phases. We find that both phases exhibit a zero-energy peak in the local density of states and spin-triplet pair correlations in the clean limit, which we attribute primarily to odd-frequency spin-triplet pairs. Disorder rapidly splits the zero-energy peak in the trivial helical phase regardless of the lengths of the superconductor and disordered normal regions. The zero-energy peak in the topological phase shows similar fragility when the superconductor region is short. However, for long superconductor regions, the zero-energy peak in the topological phase remains robust against disorder. In contrast, spin-singlet correlations are suppressed near zero energy in both phases. Our results highlight that the robustness of the zero-energy peak against scalar disorder, contingent on the superconductor region length, serves as a key indicator distinguishing trivial Andreev bound states from topological Majorana bound states.
\keywords{Majorana Zero Modes \and Andreev Bound states \and Anomalous Proximity Effect \and Odd-Frequency Pairing}
\end{abstract}

\section{Introduction}
\label{sec1}
The search for Majorana bound states (MBSs) in semiconductor-superconductor systems has spurred enormous efforts since their initial prediction \cite{Aguadoreview17,lutchyn2018majorana,zhang2019next,frolov2019quest,flensberg2021engineered,prada2019andreev}, in great part  because they characterize a topological superconducting state \cite{tanaka2012symmetry,sato2016majorana,sato2017topological,tanaka2024theory} and also due to their promise for realizing topological qubits \cite{sarma2015majorana,beenakker2019search,sola2020majorana,Marra_2022}. MBSs emerge as zero-energy edge states after a topological phase transition driven by an applied Zeeman field \cite{tanaka2024theory}, an idea that triggered an impressive amount of theoretical and experimental studies addressing the zero-energy nature of MBSs \cite{Aguadoreview17,lutchyn2018majorana,zhang2019next,frolov2019quest,flensberg2021engineered,prada2019andreev,tanaka2024theory}. Despite the apparent simple recipe, the detection of MBSs has turned out to be difficult because semiconductor-superconductor systems also host trivial zero-energy Andreev bound states (ABSs) \cite{Bagrets:PRL12,Pikulin2012A,PhysRevB.86.100503,PhysRevB.91.024514,PhysRevB.86.180503,PhysRevB.98.245407,PhysRevLett.123.117001,DasSarma2021Disorder,PhysRevB.105.144509,PhysRevB.104.L020501,PhysRevB.104.134507,marra2022majorana,PhysRevB.107.184509,baldo2023zero,PhysRevB.107.184519,PhysRevB.110.165404,PhysRevB.110.085414,PhysRevLett.132.099602,legg2025commentinasalhybriddevices,legg2024replyantipovetal,legg2025commentinterferometricsingleshotparity} that can behave as MBSs.  As a result,  identifying robust zero-energy states in semiconductor-superconductor systems is only a necessary condition for MBSs \cite{TK95,KT96,Kashiwaya_RPP,Proximityp,PhysRevLett.103.237001,PhysRevB.82.180516} and, to approach a sufficient condition, the detection protocol must also involve simultaneous measurements of other Majorana properties \cite{PhysRevB.96.085418,PhysRevB.96.205425,PhysRevB.96.201109,cayao2018andreev,PhysRevB.104.L020501,PhysRevB.105.054504,PhysRevB.110.224510,mizushima2025detecting,PhysRevB.108.205426}.

In the mid-2000s, the simultaneous analysis of multiple Majorana properties was shown to be unambiguously revealed by the so-called \emph{anomalous proximity effect} in a disordered normal metal (DN)  attached to a spin-triplet superconductor \cite{Proximityp,PhysRevB.71.094513,PhysRevB.72.140503,PhysRevLett.96.097007}. In the anomalous proximity effect, zero-energy MBSs emerging at the interface penetrate into the DN region, giving rise to a zero-energy peak (ZEP) in the local density of states (LDOS) \cite{Proximityp,PhysRevB.71.094513,PhysRevB.72.140503,PhysRevLett.99.067005,Higashitani2009}. In contrast, for  a spin-singlet $s$-wave superconductor, the conventional proximity effect is characterized by a zero-energy dip in the LDOS \cite{golubov1988theoretical,PhysRevB.54.9443}. Further studies revealed that the main actor in the anomalous proximity effect is the odd-frequency spin-triplet s-wave pairing induced into the disordered normal region  \cite{odd1,odd3,odd3b,Higashitani2009,tanaka2012symmetry,PhysRevB.87.104513,Kokkeler2022,Kokkeler2023}. The s-wave nature of this type of pairing was shown to be crucial for its robustness against scalar disorder. Moreover, the anomalous proximity effect has been also shown to originate other exotic phenomena, such as enhanced Josephson currents at low temperatures \cite{PhysRevLett.96.097007,AsanoPRB2006}, paramagnetic Meissner response \cite{PhysRevB.72.140503,Yokoyama2011}, anomalous surface impedance \cite{PhysRevLett.107.087001}, and a zero-bias voltage quantized conductance peak \cite{Proximityp,PhysRevB.71.094513,PhysRevB.94.054512,Takagi18}.  This fueled the interest in the anomalous proximity effect, leading to a series of theoretical studies \cite{tanaka2012symmetry,
Tamura2019,Suzuki2019,Mizushima2023,PhysRevB.91.174511,PhysRevB.94.054512,PhysRevB.95.214503,PhysRevB.97.174501,PhysRevB.102.140505,
NagaeFlatband2025} and culminating in the recent experimental detection of the anomalous proximity effect in CoSi$_{2}$/TiSi$_{2}$ heterostructures \cite{doi:10.1126/sciadv.abg6569,D2NR05864B,CHIU2024348}. In all cases, the anomalous proximity effect appears as a strong signature of unconventional spin-triplet superconductivity, suggesting that it can be also used to identify and understand other unconventional superconducting states. \footnote{Other systems expected to give an anomalous proximity effect are those hosting non-Hermitian effects, such as those reported in Refs.  \cite{xbj1-hfyf,PhysRevB.110.085414,PhysRevB.107.104515,JorgeEPs,PhysRevB.110.L201403,PhysRevLett.124.086801,PhysRevX.9.041015}}  Given that robust zero-energy ABSs in semiconductor-superconductor systems can behave as MBSs, and the emergent superconductivity is unconventional, it is natural to ask if such states produce an anomalous proximity effect which can improve our understanding of Majorana and Andreev physics.

\begin{figure}[!t]
    \centering
    \includegraphics[width=\linewidth]{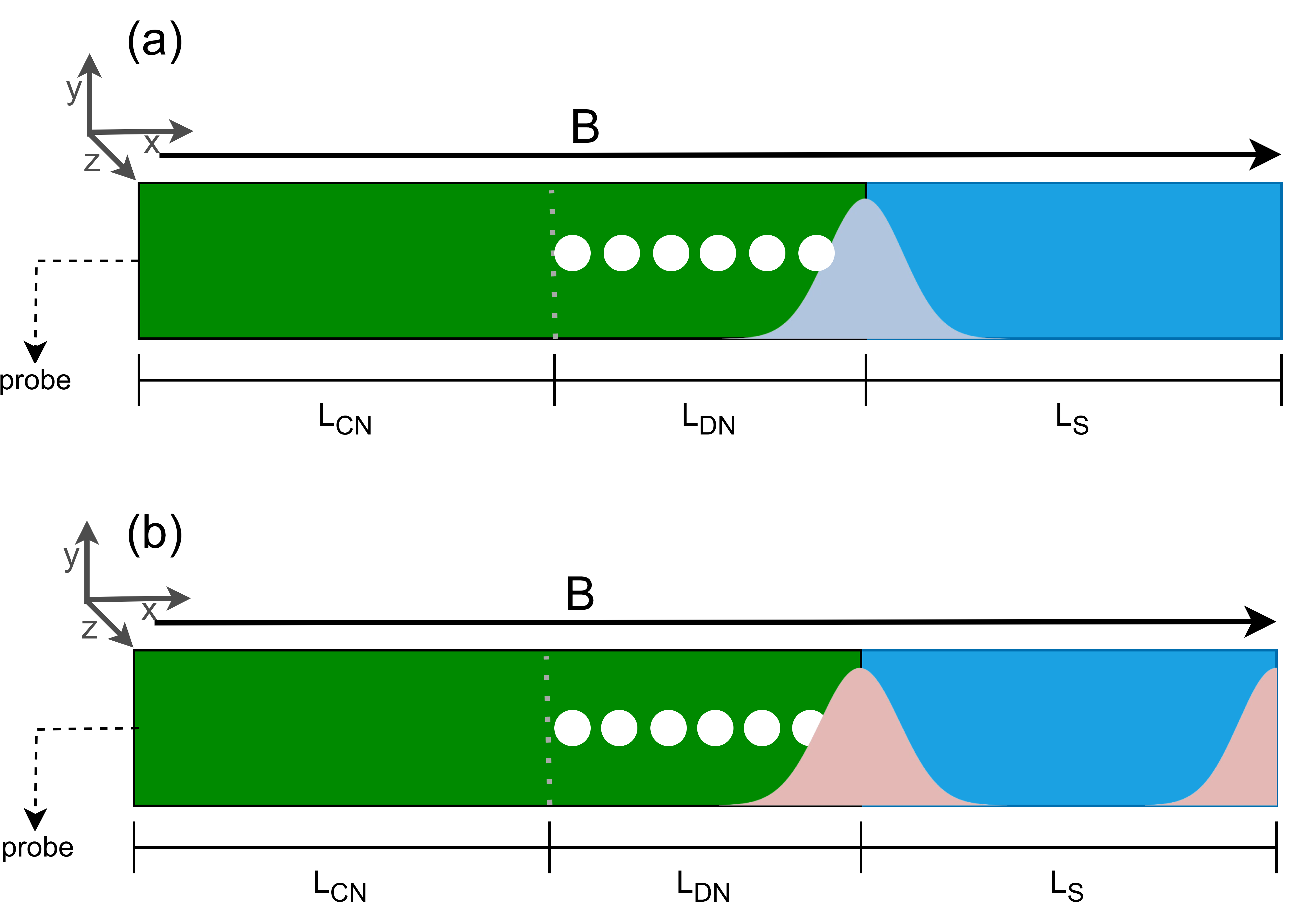}
    \caption{Schematic of the CN/DN/S junction. The left part of the system is a normal region (N, green area) with total length $L_{\rm N}$. The normal region is divided into two parts: a clean normal region (CN) with length $L_{\rm CN}$ and a disordered normal region (DN) with length $L_{\rm DN}$, with scalar onsite disorder represented as white dots.  The total length of the normal region is $L_{\rm N} = L_{\rm CN} + L_{\rm DN}$. The right part of the system is a superconducting region (S) of length $L_{\rm S}$. The system is subjected to an external Zeeman field $B$ along the $x$ direction. The chemical potential in the normal region $\mu_N$ is lower than that of the superconductor $\mu_S$. Panel (a) symbolizes the system in the trivial regime with ABSs (light blue) at the DN/S interface, whereas panel (b) symbolizes the system in the topological regime with MBSs (light red) at the ends of the S region. }
    \label{fig1}
\end{figure}

In this work, we theoretically analyze the anomalous proximity effect under the influence of both trivial zero-energy ABSs and topological MBSs in semiconductor-superconductor systems. In particular, we consider a normal-superconductor (NS) junction realized in a one-dimensional Rashba nanowire  with partially proximity-induced spin-singlet s-wave superconductivity under a uniform magnetic field along the wire, see Fig. \,\ref{fig1}. Here, the N region is further composed of a clean (C) and a disordered (D) region, which permits us to explore the anomalous proximity effect. By tuning the Zeeman field $B$, the system can be driven from a trivial helical phase, hosting zero-energy ABSs at the interface, into a topological phase hosting MBSs at the ends of the superconducting region. We investigate the signatures of the anomalous proximity effect, specifically the local density of states (LDOS) and induced spin-singlet and spin-triplet pair correlations, at the far edge of the N region using a Green's function approach. We systematically analyze how these signatures depend on the scalar disorder strength $W$ within the DN region and the lengths of the superconducting and disordered normal segments. We demonstrate that while both the trivial and topological phases exhibit a ZEP in the LDOS and odd-frequency spin-triplet pair correlations in the clean limit, the robustness of these signatures against disorder and the lengths of the S and DN regions differs significantly. In particular, we find that the ZEPs in the LDOS and odd-frequency spin-triplet pair correlations in the trivial phase are fragile against disorder, rapidly splitting and diminishing with increasing disorder, irrespective of the S and DN segment lengths. In contrast, in the topological phase, these ZEPs show a strong dependence on the S region length. While topological ZEPs are robust against disorder for long S regions, they are fragile against disorder for short S regions, mirroring the behavior in the trivial phase. Our findings thus clarify the specific conditions, particularly the necessity of a sufficiently long superconducting segment, under which anomalous proximity effect signatures can serve as reliable indicators to distinguish true topological MBSs from trivial zero-energy ABSs in such hybrid junctions.

The remainder of this paper is organized as follows:
In Sec~\ref{sec2}, we outline the theoretical model for our junction and the Green's function methodology.
In Sec.~\ref{sec3}, we analyze the low-energy spectrum.
In Sec~\ref{sec4}, we present our main results on the anomalous proximity effect, detailing the local density of states and induced pair correlations, and their robustness.
Finally, in Sec.~\ref{sec5}, we summarize our findings and conclusions.

\section{Model and Methods}
\label{sec2}
We consider a one-dimensional semiconductor nanowire with strong spin-orbit coupling\editE{. The right segment of the wire is placed on top of a conventional s-wave superconductor. Due to the proximity effect, a superconducting pairing gap $\Delta$ is induced.} We also assume that the right portion of the unproximated region hosts scalar impurities, effectively realizing a clean normal metal/disordered normal metal/superconductor (CN/DN/S) junction. The junction is subjected to an external Zeeman field along the $x$ direction. The value of the Zeeman field determines whether we are in the topological or trivial regime. We model the system using a one-dimensional tight-binding Hamiltonian on a lattice with lattice constant $a$. The Bogoliubov-de Gennes (BdG) Hamiltonian for the entire system can be written as:
\begin{equation}
    H = \sum_{i} \Psi_i^\dagger \mathcal{H}_{ii} \Psi_i + \sum_{\langle i,j \rangle} \left( \Psi_i^\dagger \mathcal{H}_{ij} \Psi_j + \text{H.c.} \right),
\end{equation}
where $\Psi_i=(c_{i\uparrow},c_{i\downarrow},c^\dagger_{i\uparrow},c^\dagger_{i\downarrow})^T$ is the Nambu spinor at site $i$, and $c_{i\sigma}^\dagger$ ($c_{i\sigma}$) creates (annihilates) an electron with spin $\sigma$ at site $i$. The summation $\langle i,j \rangle$ runs over nearest neighbors.

The onsite Hamiltonian $\mathcal{H}_{ii}$ is given by:
\begin{equation}
    \mathcal{H}_{ii} = (2t - \mu_i-v_i)\tau_z \sigma_0 + B \tau_z \sigma_x + \Delta\Theta(i-L_{\rm N}/a) \tau_y \sigma_y,
\end{equation}
and the hopping term $\mathcal{H}_{ij}$ between nearest-neighbor sites $i$ and $j=i+1$ is:
\begin{equation}
    \mathcal{H}_{ij} = -t \tau_z \sigma_0 - i \frac{\alpha}{2a} \tau_0 \sigma_z.
\end{equation}
Here, $\tau_{x,y,z}$ and $\sigma_{x,y,z}$ are Pauli matrices acting in the particle-hole and spin spaces, respectively, while $\tau_0$ and $\sigma_0$ are identity matrices. $t$ is the hopping amplitude, $\mu_i$ is the chemical potential at site $i$, $B$ is the Zeeman field, $\alpha$ is the Rashba spin-orbit coupling strength,  $\Delta$ is the proximity-induced superconducting gap, and $\Theta(x)$ is the Heaviside step function. The normal region, including both clean and disordered regions, extends from site $i=1$ to $i=L_{\rm N}/a$. The disordered region has total length $L_{\rm DN}$ and extends from site $i=(L_{\rm N}-L_{\rm DN})/a+1$ to $i=L_{\rm N}/a$. The superconducting region extends from $i=L_{\rm N}/a+1$ to $i=(L_{\rm N}+L_{\rm S})/a$ and has length $L_{\rm S}$. 
To be able to study both the trivial helical phase and topological phase within the same model, we set the chemical potential in the normal region lower than that of the superconductor region. This can be achieved, for example, via gating. Thus, the chemical potential is given by
\begin{equation}
    \mu_i =\begin{cases}
        \mu_N & i\leq L_{\rm N}/a,\\
        \mu_S & i> L_{\rm N}/a,
    \end{cases}
\end{equation}
with $\mu_N<\mu_S$. \editE{We note that our model describes a system in thermodynamic equilibrium. Hence, no net particle current flows despite the chemical potential difference.}

Disorder is introduced in the DN region via a random onsite potential $v_i$ drawn from a uniform distribution $[-W, W]$, where $W$ is the scalar onsite disorder strength. \editE{Experimentally, such disorder could be realized via ion irradiation or by controlling defects during growth \cite{nam2005disorder,holmes2025focused,kohopaa2024effect,C3NR00368J}.} Outside the DN region, the system is considered to be clean ($v_i=0$ for $i\leq (L_{\rm N}-L_{\rm DN})/a$ and for $i>L_{\rm N}/a$). 
We refer to $W \lesssim \mu_S$ as weak disorder, $W$ comparable to a few times $\mu_S$ (e.g., $W \approx (1\sim 3)\mu_S$) as moderate disorder, and $W \gg \mu_S$ as strong disorder in our discussions. \editE{We use $\mu_S$ as the reference energy scale as it sets the Fermi energy in the superconducting region. Since the topological superconductivity and the associated MBSs originate from the S region, $\mu_S$ is the natural energy scale for the quasiparticles involved in the proximity effect. Comparing the disorder potential strength $W$ to $\mu_S$ gives a measure of how disruptive the scattering is to these quasiparticles.}

\begin{figure}[!t]
    \centering
    \includegraphics[width=\linewidth]{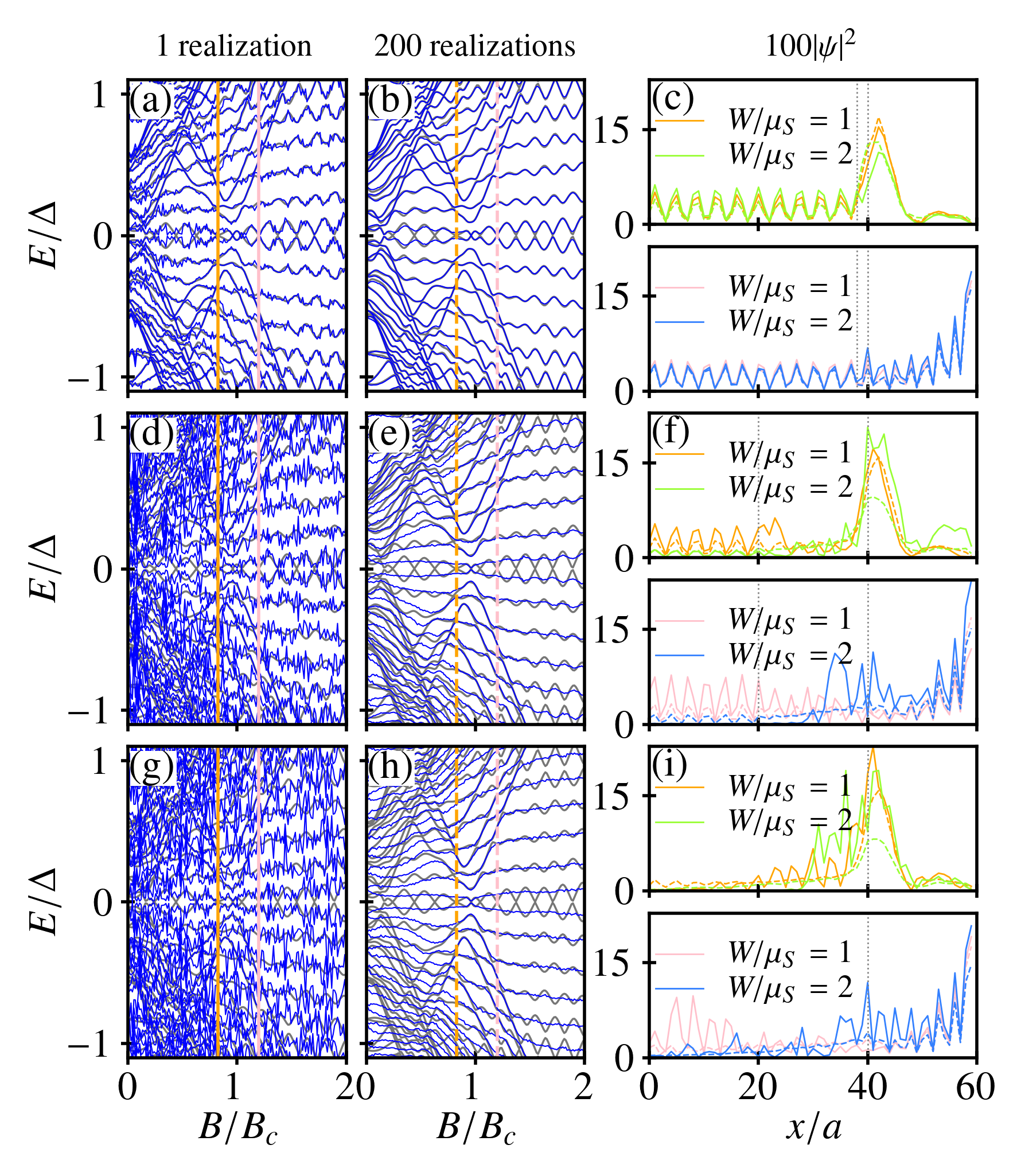}
    \caption{Low energy spectrum for CN/DN/S junction with short S region as a function of the Zeeman field $B$ at $W=\mu_S$ for $L_{\rm DN}=2a$ (a,b), $L_{\rm DN}=20a$ (d,e), and $L_{\rm DN}=40a$ (g,h). Panels (a,d,g) show the spectrum for one disorder realization, whereas panels (b,e,h) show the average spectrum over 200 disorder realizations. The light gray curves in (a,b,d,e,g,h) show the spectrum for the clean case ($W=0$). Panels (c,f,i) show the wavefunction probability density of the zero-energy states with one disorder realization (solid) and for ensemble average (dashed) with $L_{\rm DN}=2a$ (c), $L_{\rm DN}=20a$ (f), and $L_{\rm DN}=40a$ (i). The top panels in (c,f,i) show the wavefunction probability density for the trivial zero-energy ABSs at $B=0.825B_c$, whereas the bottom row shows the wavefunction probability density for the MBSs at $B=1.2B_c$. The orange and pink curves in (a,b,d,e,g,h) show the values of $B$ for which the wavefunction probability density is shown in (c,f,i). The parameters used are: $L_{\rm S}=20a$, $L_{\rm N} = 40a$, $a=50$ nm, $\mu_S=0.5$ meV, $\mu_N=0.1$ meV, $\Delta=0.25$ meV, $\alpha=20$ meV nm, $t=1$ meV, and $W=\mu_S$ unless otherwise stated.} 
    \label{fig2}
\end{figure}

The topological phase transition in the S region occurs when the condition $B =B_c = \sqrt{\Delta^2 + \mu_S^2}$ is satisfied, provided that the wire is sufficiently long \cite{tanaka2024theory}\editE{, meaning its length $L_S$ is much larger than twice the Majorana localization length ($L_S\gg 2\xi_{\rm MBS}$) to suppress the hybridization of end states}. In the topological phase ($B>B_c$), MBSs are expected to form at the ends of the S region.\footnote{Majorana states can be characterized by nontrivial topological invariants \cite{v6ht-jq3l,PhysRevB.110.165404,cheng2021fate,PhysRevB.88.020407,Kitaev_2001,PhysRevB.103.104203,PhysRevB.92.115115} as well as by unique features of their odd-frequency superconducting pairing \cite{tanaka2012symmetry,tanaka2024theory,cayao2019odd}.}

The system has another phase, the helical phase in the normal region. The helical phase starts at the helical transition point $B=\mu_N$ and extends to the region $\mu_N<B<B_c$. In this phase, the system is in a topologically trivial state and the zero-energy states are expected to be trivial ABSs \cite{PhysRevB.91.024514,PhysRevB.104.L020501}. The helical phase is characterized by the presence of a zero-energy state bound to the DN/S interface, which is a consequence of the helical nature of the system \cite{ahmed2025odd}. 

We numerically diagonalize the tight-binding Hamiltonian $H$ to obtain the energy eigenvalues $E$ and the corresponding eigenvectors $\psi_i^E = (\mathbf{u}^E_{i\uparrow}, \mathbf{u}^E_{i\downarrow}, \mathbf{v}^E_{i\uparrow}, \mathbf{v}^E_{i\downarrow})^T$.

To calculate the LDOS and pair correlations, we employ Nambu's Green's function method. The retarded (advanced) Nambu Green's function between sites $j$ and $j'$ with spin $\sigma$ and $\sigma'$ is defined as:
\begin{equation}
    \mathcal{G}^{r(a)}_{j\sigma,j'\sigma'}(E) = (E \pm i\eta - H)^{-1}_{j\sigma,j'\sigma'},
\end{equation}
where $\eta$ is a small positive number. The Nambu Green's function has the following Nambu matrix form:
\begin{equation}
    \mathcal{G}^{r(a)}_{j\sigma,j'\sigma'}(E) = \begin{pmatrix}
        G^{r(a)}_{j\sigma,j'\sigma'} & F^{r(a)}_{j\sigma,j'\sigma'} \\
        \tilde{F}^{r(a)}_{j\sigma,j'\sigma'} & \tilde{G}^{r(a)}_{j\sigma,j'\sigma'}
    \end{pmatrix},
\end{equation}
where $G^{r(a)}_{j\sigma,j'\sigma'}$ is the normal Green's function, $F^{r(a)}_{j\sigma,j'\sigma'}$ is the anomalous Green's function, and $\tilde{F}^{r(a)}_{j\sigma,j'\sigma'}$ and $\tilde{G}^{r(a)}_{j\sigma,j'\sigma'}$ are their corresponding particle-hole conjugates. The LDOS can be calculated from the normal Green's function as:
\begin{equation}
    \rho(E,i) = -\frac{1}{\pi} \text{Im}  \left[ Tr( \mathcal{G}^{r}_{ii}(E)) \right],
\end{equation}
where the trace is taken over the spin and particle-hole indices. 
The pair correlations can be extracted from the anomalous Green's function by rewriting it in the following form:
\begin{equation}
    \hat{F} = (d_s + \Vec{d}\cdot \Vec{\sigma}) (i\sigma_y),
\end{equation}
where $\hat{}$ denotes a matrix in spin space, $d_s$ is the spin-singlet component and $\Vec{d}=(d_x,d_y,d_z)^T$ are the spin-triplet components of the pair correlations, and we have dropped spatial indices for clarity. \editE{While the topological superconductivity in the clean S region is known to be of an effective {\it p}-wave nature, pair correlations with non-zero angular momentum are fragile against the scalar disorder present in the DN region.} It has been shown that only local \editE{{\it s}-wave} pairing correlations can survive in the presence of disorder and hence, we only consider local pairing correlations in this paper \cite{odd1}. The local pair correlations can be calculated as:
\begin{align}
    d_s(i,E) &=  \frac{F_{i\uparrow,i\downarrow}(E) -F_{i\downarrow,i\uparrow}(E)}{2}, \\
    d_x(i,E) &=  \frac{F_{i\downarrow,i\downarrow}(E) -F_{i\uparrow,i\uparrow}(E)}{2}, \\
    d_y(i,E) &=  i\frac{F_{i\downarrow,i\uparrow}(E) +F_{i\uparrow,i\downarrow}(E)}{2}, \\
    d_z(i,E) &=  \frac{F_{i\uparrow,i\downarrow}(E) +F_{i\downarrow,i\uparrow}(E)}{2}.
\end{align}
Due to fermionic statistics, pair amplitudes must be overall antisymmetric under particle exchange. For local ($s$-wave) correlations, spatial symmetry is even. Therefore, the spin-singlet component $d_s$ must be even in frequency, while the spin-triplet components $d_x$, $d_y$, and $d_z$ must be odd in frequency \cite{RevModPhys.77.1321,tanaka2012symmetry,tanaka2024theory,
Balatsky2017,cayao2019odd,triola2020role}.
Having outlined the model and methods, we now list the parameters used in this work. We use realistic parameters for our model based on InSb and InAs nanowires \cite{lutchyn2018majorana}. For numerical convenience, we choose a large lattice constant $a=50$ nm, which is five times larger than the value used in the literature. This allows us to use a smaller number of lattice sites in our calculations. The parameters used are: the hopping amplitude $t=1$ meV, the chemical potential in the superconductor $\mu_S=0.5$ meV, the chemical potential in the normal region $\mu_N=0.1$ meV, the superconducting gap $\Delta=0.25$ meV, and the spin-orbit coupling strength $\alpha=20$ meV nm. The disorder strength $W$ is varied between $0$ and $5\mu_S$. The length of the S region $L_{\rm S}$ is varied between $L_{\rm S}=20a$ and  $L_{\rm S}=100a$), while the total length of the normal region is fixed at $L_{\rm N} = 40a$. The length of the disordered region $L_{\rm DN}$ is varied ($L_{DN}=2a, 20a, 40a$), which determines the length of the clean normal region $L_{\rm CN} = L_{\rm N} - L_{\rm DN}$. \editE{The total length of the normal region is fixed to ensure constant confinement conditions, which is necessary for the formation of confinement-induced helical Andreev bound states in the trivial regime \cite{ahmed2025odd}. By varying $L_{\rm DN}$ within this fixed length, we can isolate the effect of disorder from the effects of changing confinement.} The critical Zeeman field for the topological transition is $B_c = \sqrt{\Delta^2+\mu_S^2} \approx 0.56$ meV. The broadening parameter is set to $\eta = 10^{-3}$. Results involving disorder are averaged over 200 independent of disorder realizations, unless otherwise specified.
\section{Zeeman dependent spectrum and low-energy states}
\label{sec3}
In this section, we analyze the low-energy spectrum of the CN/DN/S junction, focusing on the behavior of the lowest-energy states under disorder. We first examine the case of a short S region to establish the spectral features of the trivial and topological phases. We then investigate how these features, particularly the lowest postive energy level, depend on the S region length and increasing disorder strength.
\subsection{Low-Energy Spectrum for Short S Regions}
\label{sec3a}
\begin{figure*}[!t]
    \centering
    \includegraphics[width=\linewidth]{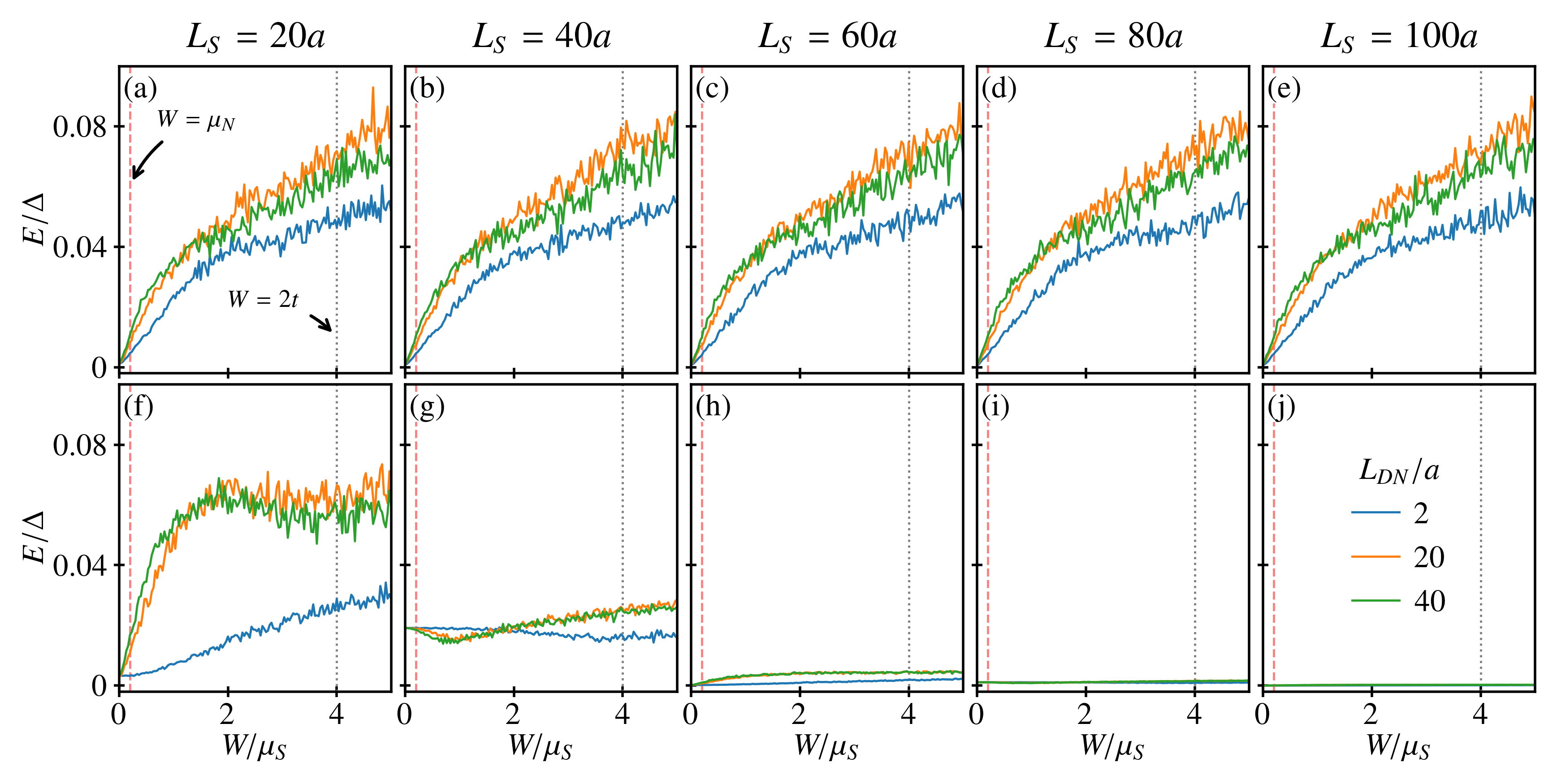}
    \caption{Lowest positive energy level $E$ as a function of scalar disorder strength $W$ in the trivial $B=0.825B_c$ (a,b,c,d,e) and topological $B=1.2B_c$ (f,g,h,i,j) phases for $L_{\rm S}=20a$ (a,f), $L_{\rm S}=40a$ (b,g), $L_{\rm S}=60a$ (c,h), $L_{\rm S}=80a$ (d,i), and $L_{\rm S}=100a$ (e,j). The parameters used are: $L_{\rm N} = 40a$, $a=50$ nm, $\mu_S=0.5$ meV, $\mu_N=0.1$ meV, $\Delta=0.25$ meV, $\alpha=20$ meV nm, $t=1$ meV, $B=0.825B_c$ for the trivial phase and $B=1.2B_c$ for the topological phase.}
    \label{fig3}
\end{figure*}
We begin by examining the low-energy spectrum of the CN/DN/S junction as a function of the Zeeman field $B$. Fig.\,\ref{fig2} shows the low-energy spectrum as well as the wavefunction probability amplitude of the zero-energy bound states for a short superconducting region ($L_S=20a$) and varying lengths of the disordered normal region ($L_{DN}$), considering both single disorder realizations ($W=\mu_S$) and ensemble averages.

In the clean limit $W=0$ (light gray curves in Fig.\,\ref{fig2} (a,b,d,e,g,h)), the spectrum exhibits the expected behavior: the bulk superconducting gap closes and reopens around the critical field $B_c =\sqrt{\Delta^2+\mu_S^2}\approx 0.56$ meV, marking the topological phase transition. For $B>B_c$, zero-energy MBSs appear at the ends of the S region. Due to the finite length $L_S=20a$, these MBSs hybridize, leading to energy eigenvalues that oscillate around $E=0$ as $B$ increases. The spectrum in the clean limit has another gap closing at $B = \mu_N$ purely due to confinement and helicity in the N region\cite{ahmed2025odd}, signaling that the system has entered into a trivial helical regime. In the helical regime  ($\mu_N < B < B_c$), ABSs emerge, and their energies oscillate around zero in a similar fashion to the Majorana oscillations in the topological regime.

\begin{figure*}[!t]
    \centering
    \includegraphics[width=\linewidth]{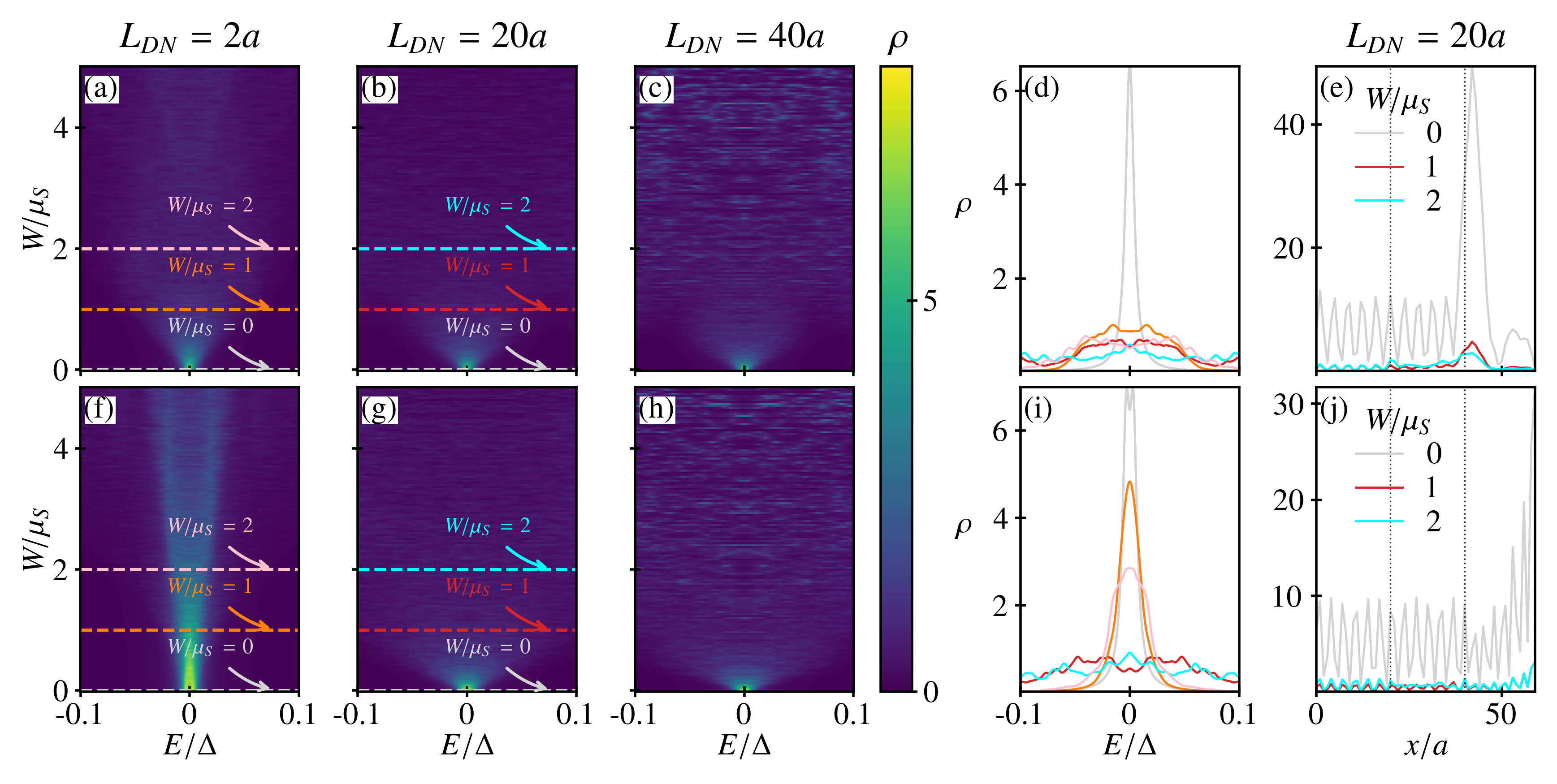}
    \caption{\editE{Ensemble-averaged} LDOS for short S region ($L_{\rm S}=20a$) junctions evaluated at the leftmost site of the N region ($i=1$) as a function of energy $E$ and scalar disorder strength $W$  with $L_{\rm DN}=2a$ (a,f), $L_{\rm DN}=20a$ (b,g), and $L_{\rm DN}=40a$ (c,h). Panels (d,i) show line cuts of the LDOS in (a,f,b,g) for $W/\mu_S=0,1,2$. Panels (e,j) show the spatial profile of the LDOS at $E \approx 0$ for $W/\mu_S=0,1,2$ with $L_{\rm DN}=20a$. Top row (a,b,c,d,e): trivial phase ($B=0.825B_c$). Bottom row (f,g,h,i,j): topological phase ($B=1.2B_c$). Other parameters used are: $L_{\rm S}=20a$, $L_{\rm N} = 40a$, $a=50$ nm, $\mu_S=0.5$ meV, $\mu_N=0.1$ meV, $\Delta=0.25$ meV, $\alpha=20$ meV nm, $t=1$ meV, and $\eta=0.001$ meV.}
    \label{fig4}
\end{figure*}

\begin{figure*}[!t]
    \centering
    \includegraphics[width=\linewidth]{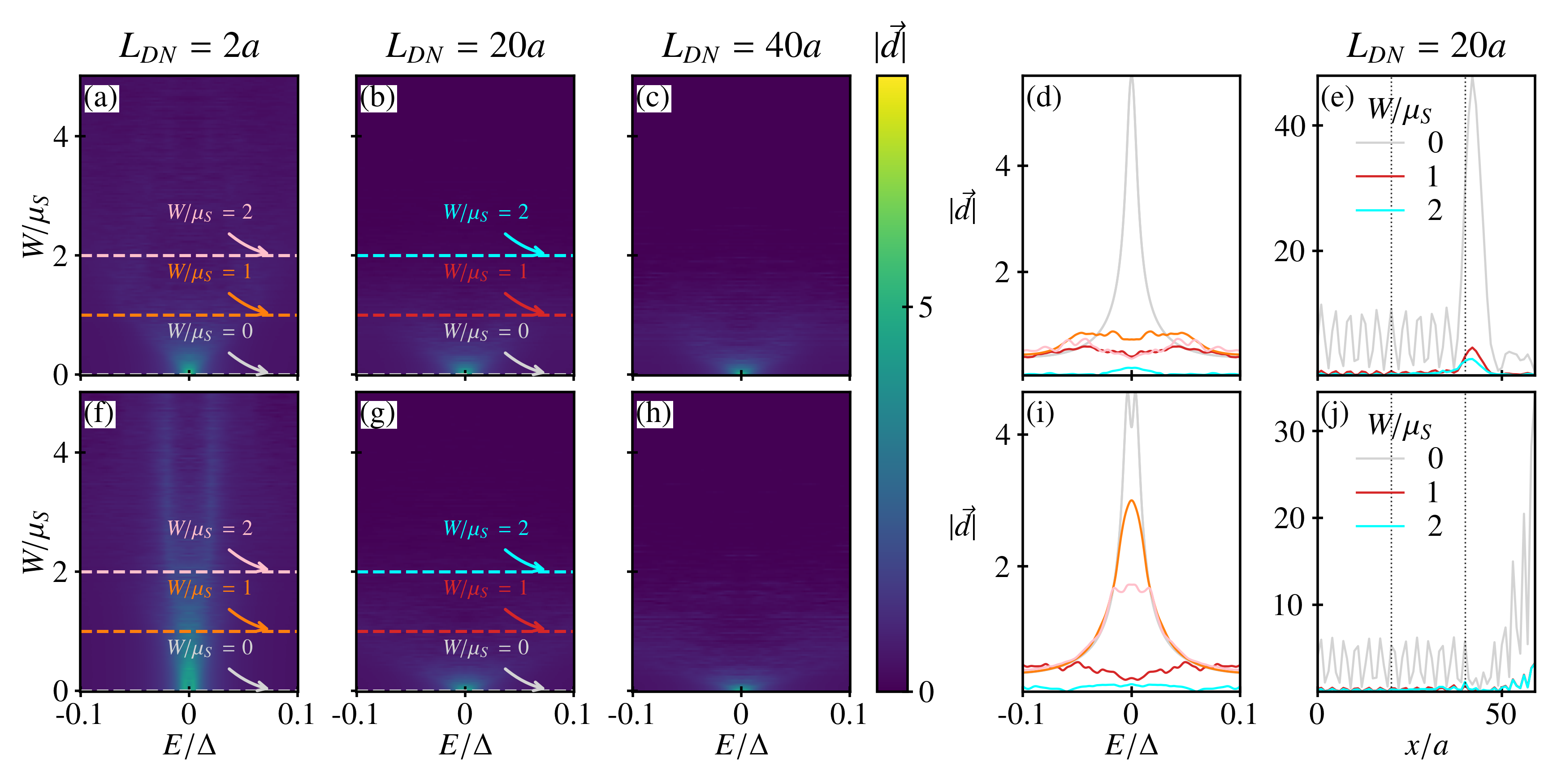}
    \caption{\editE{Ensemble-averaged} Absolute value of the spin-triplet pair correlations $|\Vec{d}|$ for short S region ($L_{\rm S}=20a$) junctions evaluated at the leftmost site of the N region ($i=1$) as a function of energy $E$ and scalar disorder strength $W$  with $L_{\rm DN}=2a$ (a,f), $L_{\rm DN}=20a$ (b,g), and $L_{\rm DN}=40a$ (c,h). Panels (d,i) show line cuts of the spin-triplet pair correlations in (a,f,b,g) for $W/\mu_S=0,1,2$. Panels (e,j) show the spatial profile of the spin-triplet pair correlations at $E\approx 0$ for $W/\mu_S=0,1,2$ with $L_{\rm DN}=20a$. Top row (a,b,c,d,e): trivial phase ($B=0.825B_c$). Bottom row (f,g,h,i,j): topological phase ($B=1.2B_c$). Other parameters used are: $L_{\rm S}=20a$, $L_{\rm N} = 40a$, $a=50$ nm, $\mu_S=0.5$ meV, $\mu_N=0.1$ meV, $\Delta=0.25$ meV, $\alpha=20$ meV nm, $t=1$ meV, and $\eta=0.001$ meV.}
    \label{fig5}
\end{figure*}
\begin{figure*}[!t]
    \centering
    \includegraphics[width=\linewidth]{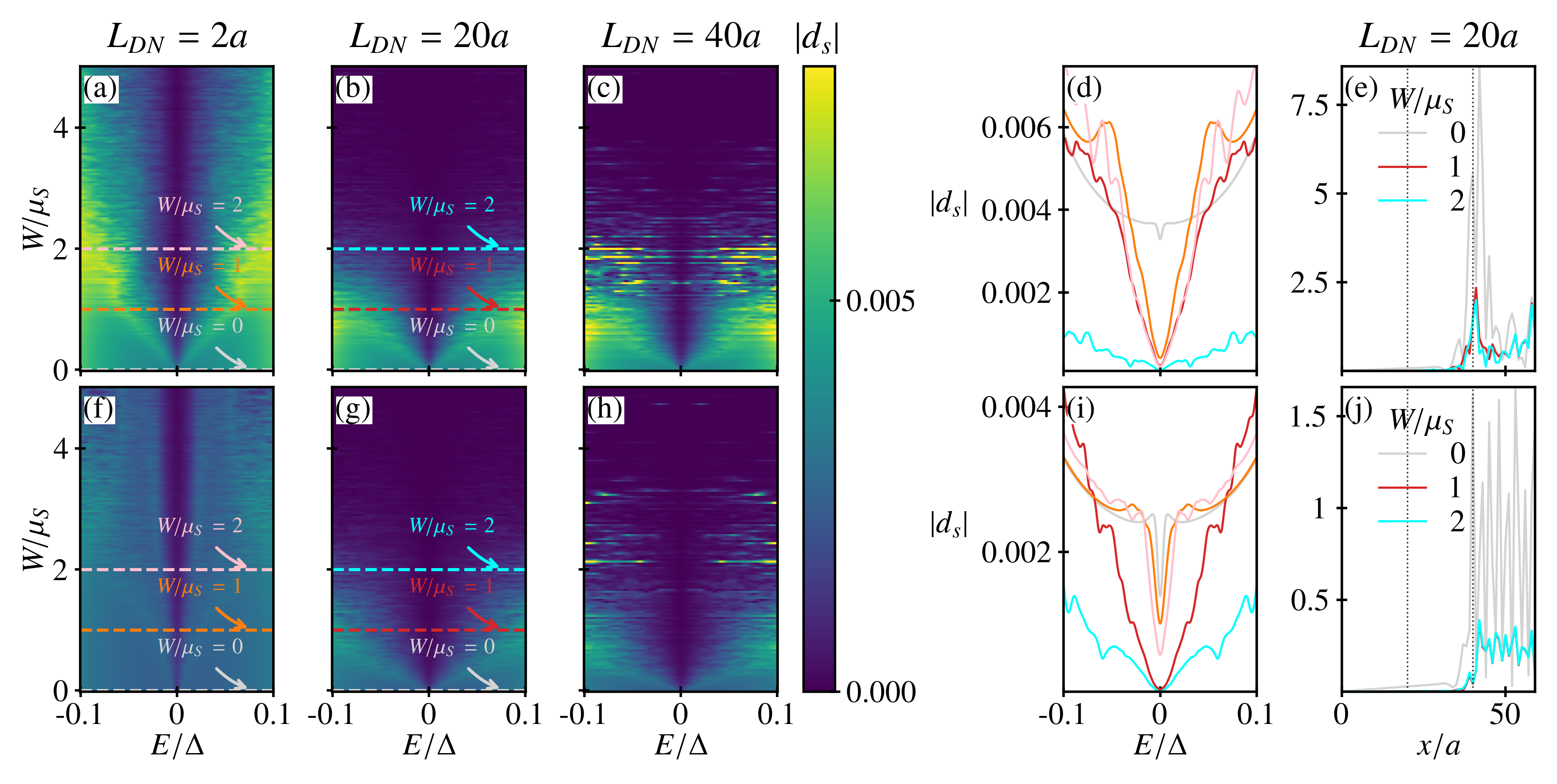}
    \caption{\editE{Ensemble-averaged} Absolute value of the spin-singlet pair correlation for short S region ($L_{\rm S}=20a$) junctions evaluated at the leftmost site of the N region ($i=1$) as a function of energy $E$ and scalar disorder strength $W$  with $L_{\rm DN}=2a$ (a,f), $L_{\rm DN}=20a$ (b,g), and $L_{\rm DN}=40a$ (c,h). Panels (d,i) show line cuts of the spin-singlet pair correlations in (a,f,b,g) for $W/\mu_S=0,1,2$. Panels (e,j) show the spatial profile of the spin-singlet pair correlations at $E\approx 0$ for $W/\mu_S=0,1,2$ with $L_{\rm DN}=20a$. Top row (a,b,c,d,e): trivial phase ($B=0.825B_c$). Bottom row (f,g,h,i,j): topological phase ($B=1.2B_c$). Other parameters used are: $L_{\rm S}=20a$, $L_{\rm N} = 40a$, $a=50$ nm, $\mu_S=0.5$ meV, $\mu_N=0.1$ meV, $\Delta=0.25$ meV, $\alpha=20$ meV nm, $t=1$ meV, and $\eta=0.001$ meV.}
    \label{fig6}
\end{figure*}

Introducing disorder ($W=\mu_S$) in the DN region significantly impacts the low-energy spectrum, as shown by the blue curves in Fig.\,\ref{fig2}(a,b,d,e,g,h). For a single disorder realization (Fig.\,\ref{fig2}(a,d,g)), the spectrum becomes much denser, and energy levels fluctuate considerably with $B$. The energy levels frequently approach or cross zero, even within the trivial regime ($B<B_c$). This is particularly true for longer DN regions. After ensemble averaging (Fig.\,\ref{fig2}(b,e,h)), these rapid fluctuations are smoothed out. We see that when the DN region is short (Fig.\,\ref{fig2}(b)), the spectrum retains its shape indicating that disorder has minimal effect on the low energy spectrum. However, for longer DN regions (Fig.\,\ref{fig2}(e,h)), both the trivial and topological zero modes split and acquire finite energy. On the other hand, we notice that when the bound states are not exactly at zero energy, disorder can, in fact, decrease the energy of the bound state (compare the gray and blue curves near oscillation maxima in Fig.\,\ref{fig2}(e,h)). Interestingly, we see that for long DN regions, the lowest positive energy level in both the trivial and topological phases disperses linearly with the Zeeman field $B$ (see Fig.\,\ref{fig2}(e,h)). This is in contrast to the oscillatory behavior observed for short DN regions (see Fig.\,\ref{fig2}(b)).

The spatial profile of the zero-energy bound states is shown in Fig.\,\ref{fig2}(c,f,i) for a single disorder realization (solid lines) and for the ensemble average (dashed lines) for $W=\mu_S$ and $W=2\mu_S$. The top panels show the wavefunction probability density for the trivial zero-energy ABSs at $B=0.825B_c$, while the bottom panels show the wavefunction probability density for the MBSs at $B=1.2B_c$. We see that for short DN regions (Fig.\,\ref{fig2}(c)), disorder has little effect on the wavefunction profile of the zero-energy states. In the helical phase, the wavefunction is localized at the DN/S interface with an oscillatory profile in the N region. In the topological phase, the wavefunction is localized at the ends of the S region, with the MBSs having a similar oscillatory profile across the entire N region, in agreement with the clean limit \cite{ahmed2025odd}. For longer DN regions (Fig.\,\ref{fig2}(f,i)), the wavefunction profile of the zero-energy states is significantly affected by disorder. While the wavefunctions of both the trivial and topological zero-energy states remain localized at the NS interface and the ends of the S region, respectively, the oscillatory profile in the entire N region is lost. Instead, the wavefunction profile is localized at the DN/S interface with an exponential decay into the DN region. This decay continues even in the clean region. The wavefunction profile of the zero-energy states is also affected by the disorder strength $W$. As disorder increases, the wavefunction profile becomes more localized at the DN/S interface and decays more rapidly into the N region. This is particularly evident in the topological phase, where the wavefunction of the MBSs becomes more localized at the DN/S interface rather than leaking into the N region, as observed in the clean limit.

\subsection{Dependence on S Region Length}
\label{sec3b}
We now turn our attention to the effect of disorder and the length of the S region on the lowest positive energy level in the trivial and topological phases. In Fig.\,\ref{fig3}, we show the lowest positive energy level in the trivial (a,b,c,d,e) and topological (f,g,h,i,j) phases as a function of disorder strength $W$ for different lengths of the S region and different lengths of DN region. Here, we chose $B =0.825B_c$ for the trivial phase and $B=1.2B_c$ for the topological phase, which correspond to the values of $B$ for which the lowest positive energy level is at the zero energy crossing in the clean limit for $L_S=20a$, as shown in Fig. \,\ref{fig2}(a,b). 
We note that for $B=1.2B_{c}$, the energy level does not start at zero for $L_{\rm S}=40a$ and $L_{\rm S}=80a$ in the topological phase, as shown in Fig.\,\ref{fig3}(g,i). 

We observe that when the superconductor is short ($L_{\rm S}=20a$), the lowest positive energy level in both the trivial and topological phases increases with disorder strength $W$ (see Fig. \,\ref{fig3}(a,f)). Initially, for weak disorder, the energy level increases linearly with $W$ in both phases. However, for strong disorder, the energy level in the trivial phase continues to increase linearly but at a slower rate. Meanwhile, in the topological phase, the lowest positive energy level saturates and remains almost constant for strong disorder, developing a plateau. This indicates that the topological phase is more robust against disorder compared to the topological trivial phase. As the length of the S region increases, the lowest positive energy level in the trivial phase as a function of disorder strength $W$ remains relatively unchanged, independent of $L_{\rm S}$ (see Fig.\,\ref{fig3}(b,c,d,e)). This is in contrast to the topological phase, where the effect of disorder is significantly reduced. For long S regions ($L_{\rm S}=80a$ and $L_{\rm S}=100a$), the lowest positive energy level in the topological phase remains at zero energy regardless of the disorder strength (see Fig.\,\ref{fig3}(i,j)). This indicates that the MBSs are robust against disorder and can survive even in the presence of the strong disorder. Interestingly, when the lowest positive energy level in the topological phase is not exactly at zero energy in clean limit (see Fig.\,\ref{fig3}(g)), the energy initially decreases with disorder strength $W$ before increasing again and eventually saturating and developing a plateau. This behavior is also observed in the trivial phase when we consider the lowest positive energy level away from zero in the clean limit; the only difference is that the energy in the trivial phase continues to increase with disorder strength $W$ without developing a plateau. However, we do not present this in this paper.

The effect of changing the length of the DN region on the lowest positive energy level is worth noting. We see that the effect of $L_{\rm DN}$ on the lowest positive energy level in the trivial phase is negligible. However, in the topological phase, the lowest positive energy level is significantly affected by the length of the DN region. For short DN regions ($L_{\rm DN}=2a$), the lowest positive energy level in the topological phase is significantly lower than that of longer DN regions and disperses linearly for larger disorder strengths. 

Thus, we conclude that the lowest positive energy level in the trivial phase is very sensitive to disorder while being insensitive to the lengths of the S and DN regions. On the contrary, the lowest positive energy level in the topological phase is very sensitive to the length of the S and DN regions, while being robust to disorder, especially for long S regions. This robustness against disorder is a hallmark of MBSs. However, we also see that for short S regions, the trivial and topological phases exhibit similar behavior in the presence of disorder, making it difficult to distinguish between them. This property is also satisfied for long DN regions, where the lowest positive energy level in both phases behaves similarly. However, for short DN regions, the trivial and topological phases can be distinguished by their different behaviors in the presence of disorder. This is particularly evident in the LDOS and pair correlations, which we discuss below.

\begin{figure*}[!t]
    \centering
    \includegraphics[width=\linewidth]{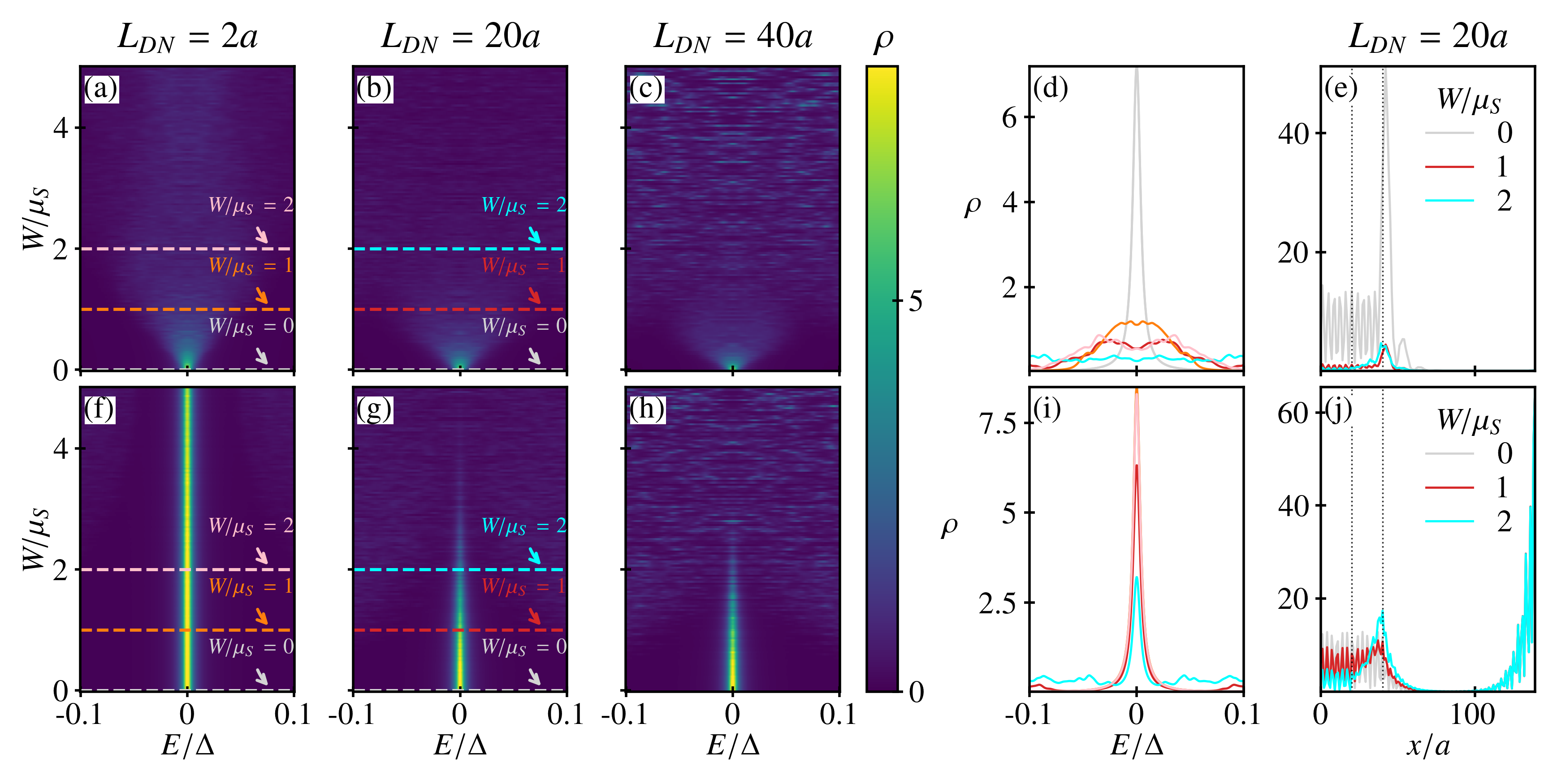}
    \caption{\editE{Ensemble-averaged} LDOS for long S region ($L_{\rm S}=100a$) junctions evaluated at the leftmost site of the N region ($i=1$) as a function of energy $E$ and scalar disorder strength $W$ with $L_{\rm DN}=2a$ (a,f), $L_{\rm DN}=20a$ (b,g), and $L_{\rm DN}=40a$ (c,h). Panels (d,i) show line cuts of the LDOS in (a,f,b,g) for $W/\mu_S=0,1,2$. Panels (e,j) show the spatial profile of the LDOS at $E\approx 0$ for $W/\mu_S=0,1,2$ with $L_{\rm DN}=20a$. Top row (a,b,c,d,e): trivial phase ($B=0.825B_c$). Bottom row (f,g,h,i,j): topological phase ($B=1.2B_c$). Other parameters used are: $L_{\rm S}=100a$, $L_{\rm N} = 40a$, $a=50$ nm, $\mu_S=0.5$ meV, $\mu_N=0.1$ meV, $\Delta=0.25$ meV, $\alpha=20$ meV nm, $t=1$ meV, and $\eta=0.001$ meV.}
    \label{fig7}
\end{figure*}
\begin{figure*}[!t]
    \centering
    \includegraphics[width=\linewidth]{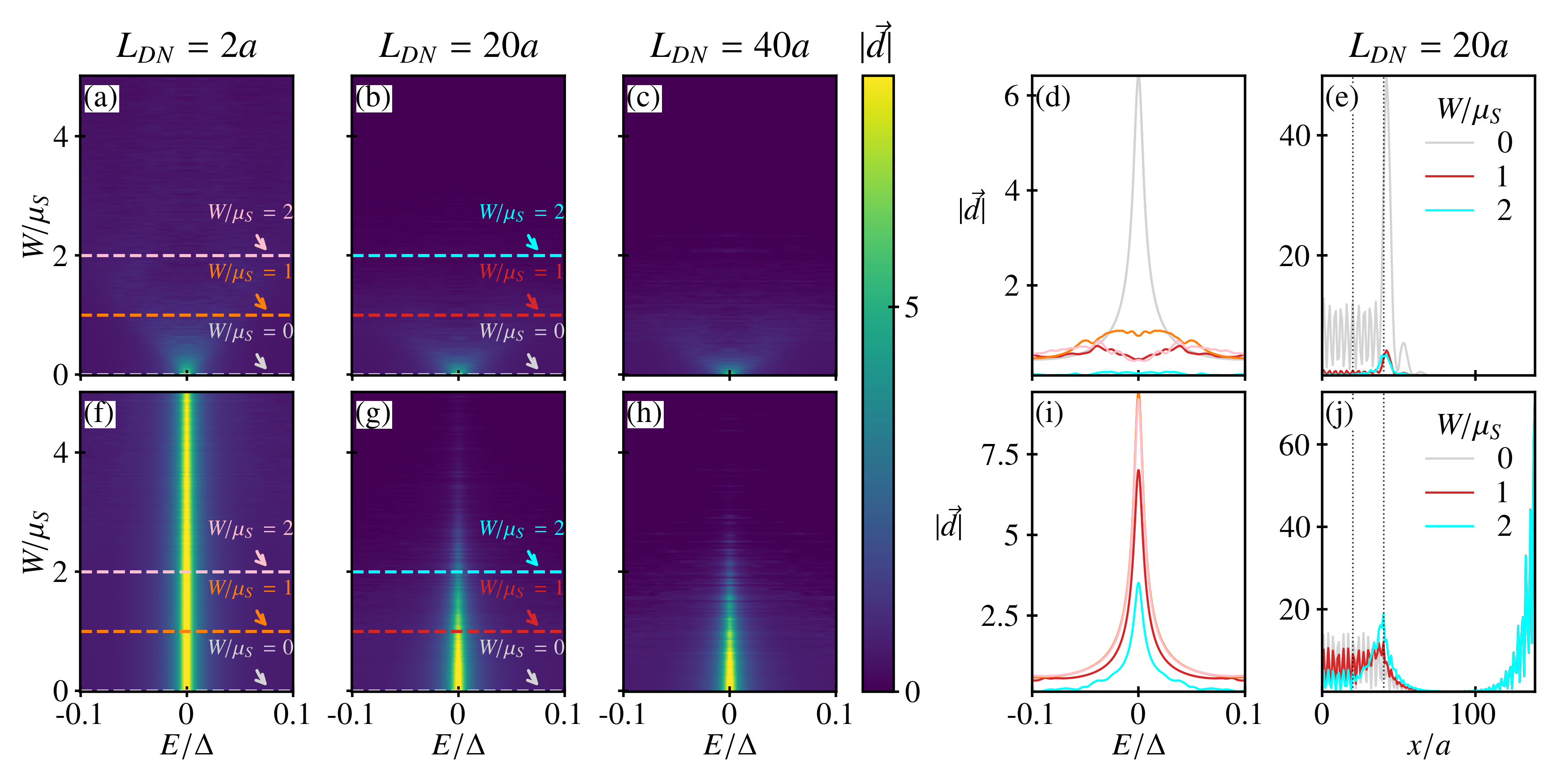}
    \caption{\editE{Ensemble-averaged} Absolute value of the spin-triplet pair correlations $|\Vec{d}|$ for long S region ($L_{\rm S}=100a$) junctions evaluated at the leftmost site of the N region ($i=1$) as a function of energy $E$ and scalar disorder strength $W$ with $L_{\rm DN}=2a$ (a,f), $L_{\rm DN}=20a$ (b,g), and $L_{\rm DN}=40a$ (c,h). Panels (d,i) show line cuts of the spin-triplet pair correlations in (a,f,b,g) for $W/\mu_S=0,1,2$. Panels (e,j) show the spatial profile of the spin-triplet pair correlations at $E\approx 0$ for $W/\mu_S=0,1,2$ with $L_{\rm DN}=20a$. Top row (a,b,c,d,e): trivial phase ($B=0.825B_c$). Bottom row (f,g,h,i,j): topological phase ($B=1.2B_c$). Other parameters used are: $L_{\rm S}=100a$, $L_{\rm N} = 40a$, $a=50$ nm, $\mu_S=0.5$ meV, $\mu_N=0.1$ meV, $\Delta=0.25$ meV, $\alpha=20$ meV nm, $t=1$ meV, and $\eta=0.001$ meV.}
    \label{fig8}
\end{figure*}
\begin{figure*}[!t]
    \centering
    \includegraphics[width=\linewidth]{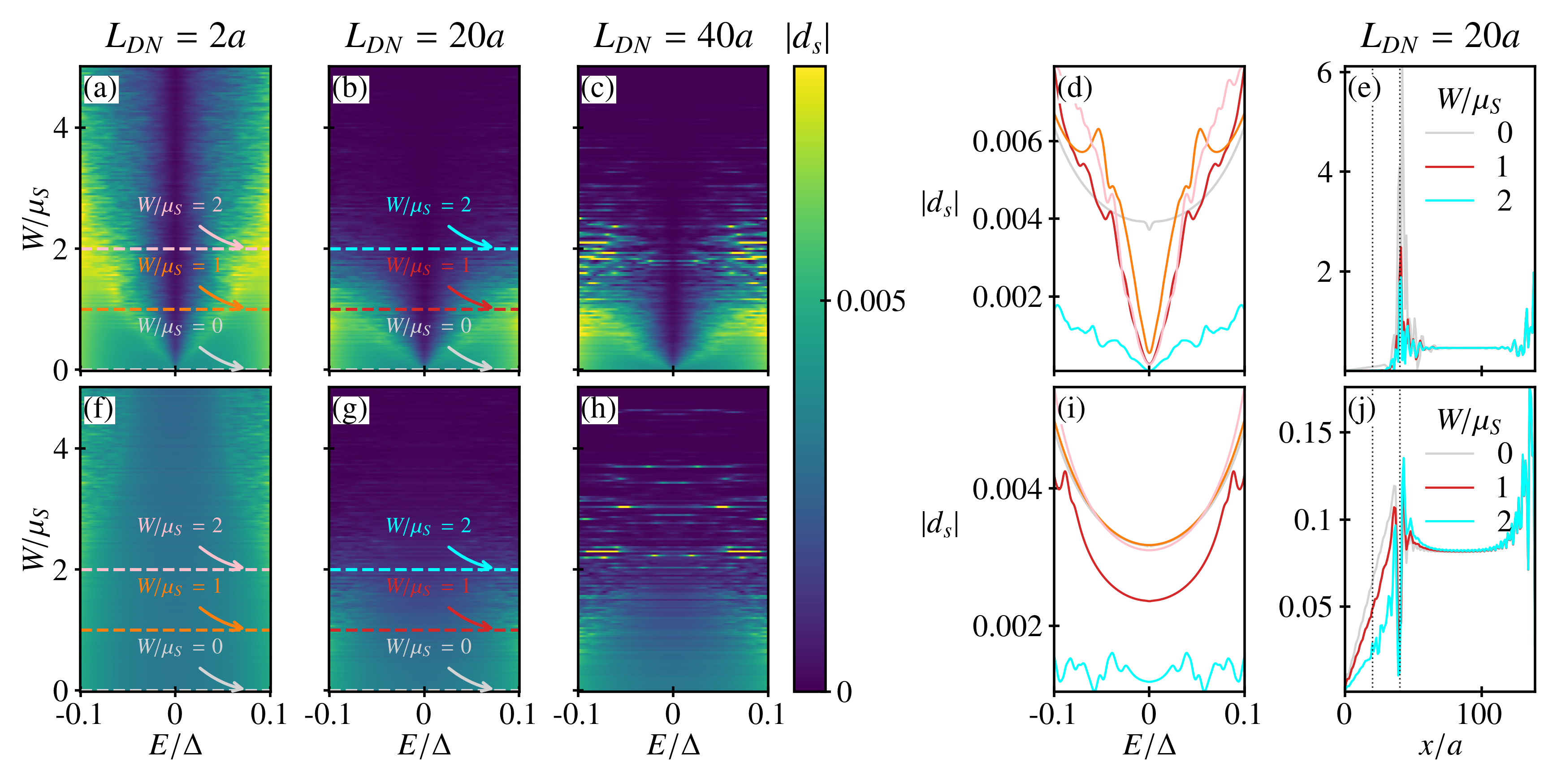}
    \caption{\editE{Ensemble-averaged} Absolute value of the spin-singlet pair correlation for long S region ($L_{\rm S}=100a$) junctions evaluated at the leftmost site of the N region ($i=1$) as a function of energy $E$ and scalar disorder strength $W$ with $L_{\rm DN}=2a$ (a,f), $L_{\rm DN}=20a$ (b,g), and $L_{\rm DN}=40a$ (c,h). Panels (d,i) show line cuts of the spin-singlet pair correlations in (a,f,b,g) for $W/\mu_S=0,1,2$. Panels (e,j) show the spatial profile of the spin-singlet pair correlations at $E\approx 0$ for $W/\mu_S=0,1,2$ with $L_{\rm DN}=20a$. Top row (a,b,c,d,e): trivial phase ($B=0.825B_c$). Bottom row (f,g,h,i,j): topological phase ($B=1.2B_c$). Other parameters used are: $L_{\rm S}=100a$, $L_{\rm N} = 40a$, $a=50$ nm, $\mu_S=0.5$ meV, $\mu_N=0.1$ meV, $\Delta=0.25$ meV, $\alpha=20$ meV nm, $t=1$ meV, and $\eta=0.001$ meV.}
    \label{fig9}
\end{figure*}
\section{LDOS and proximity-induced pair correlations}
\label{sec4}
Having discussed the low-energy spectrum and the nature of the zero-energy states in the CN/DN/S junction, we now turn our attention to the LDOS and pair correlations. Given that the trivial and topological phases have similar low-energy spectra in junctions with short S regions, we first focus on the case of short S regions ($L_S=20a$) and discuss the effect of disorder and the length of the DN region on the LDOS and pair correlations.

\editE{\subsection{Physical picture: Disorder and the role of chiral symmetry}}
\label{sec4 physical picture}
\editE{Before we present our numerical results, it is useful to establish the physical mechanism governing the stability of the ZEP against disorder. The key to this protection is the chiral symmetry of the BdG Hamiltonian \cite{PhysRevB.94.054512,PhysRevB.95.214503,PhysRevB.97.174501}. Generally speaking, for any chiral symmetric systems, any zero-energy state (ZES) is necessarily an eigenstate of the chiral operator $\Gamma$, and has a definite chirality ($\gamma =\pm1$). Meanwhile, any nonzero energy state must be an equal superposition of states with opposite chiralities. This is a direct consequence of the the chiral symmetry condition ($\Gamma H \Gamma = - H$). If chiral symmetry preserving disorder (e.g. scalar disorder) is introduced into the system, disorder can only move a ZES to a finite energy if it can couple it with another ZES of the opposite chirality.  The number of robust ZES is therefore given by the index $N_{ZES} = |N_+ - N_-|$, which represents the imbalance between the number of states with positive ($N_+$) and negative ($N_-$) chirality \cite{PhysRevB.94.054512}.}

\editE{This framework provides a clear, microscopic picture for our results that hinges on the spatial properties of the chiral partner states. For the topological MBSs, the two states with opposite chirality are the two MBSs themselves, localized at opposite ends of the superconducting region. In a long S-region, these two states are spatially well-separated. A local perturbation like scalar disorder in the DN region cannot simultaneously couple these distant states. Therefore, the MBS at the DN/S interface remains robustly pinned at zero energy. The MBSs in a short S-region are fragile for the same reason. Mainly, their wavefunctions overlap significantly, allowing local disorder to couple them and left them away from zero energy.}

\editE{For the trivial ABS, the situation is drastically different. The ABS itself is a superposition of two spatially overlapping eigenstates of the chiral operator with opposite chirality. One eigenstate is localized precisely at the DN/S interface and is responsible for the localized behavior observed in the top panels of Fig.\,\ref{fig2}(c,f,i), while the other is a more delocalized, particle-in-a-box-like state within the normal region and is responsible for the oscillatory behavior of the wavefunction in the top panels of Fig.\,\ref{fig2}(c,f,i). Because these two opposite-chirality components coexist in the same location, local disorder can easily couple them, lifting the degeneracy and pushing the ABS to a finite energy. This explains its fragility.}

\editE{Crucially, the ZEP observed in the LDOS probed at the leftmost site of the junction should reflect the physics discussed above. In particular, we should expect fragile ZEPs in the trivial phase and in the topological phase of junctions with short S region. In contrast, the ZEP should be robust in the topological phase when the MBSs are well-separated which occurs in junctions with long S regions.}

\subsection{Anomalous Proximity Effect for Short S Regions}
\label{sec4a}
In Fig.\,\ref{fig4}, we show the LDOS at the leftmost site of the N region ($i=1$) as a function of energy $E$ and disorder strength $W$ for different lengths of the DN region ($L_{DN}$). The top row (a,b,c,d,e) shows the LDOS for the trivial phase, while the bottom row (f,g,h,i,j) shows the LDOS for the topological phase. We see that in the absence of disorder, both the trivial and topological phases show a ZEP in the LDOS. As disorder is introduced into the system, the ZEP splits very quickly in the trivial phase, developing a fan-shaped structure regardless of the length of the disordered region, as shown in Fig.\,\ref{fig4}(a,b,c). Interestingly, the ZEP in the topological phase also splits under disorder, particularly for longer DN regions (see Fig.\,\ref{fig4}(f,g,h)). Surprisingly, we see that for long DN regions, the ZEP in the topological phase displays the same fan structure as in the trivial phase with the same size and shape. This indicates that the LDOS is not a reliable probe for distinguishing between the trivial and topological phases in the presence of disorder for long DN regions. However, for short DN regions, the ZEP in the topological phase can survive for weak disorder and acquire a small energy splitting (see Fig.\,\ref{fig4}(f)). Thus, for short S regions, it is possible to distinguish between the trivial and topological phases using anomalous proximity effect only when the disordered region is very short. Here, we remind the reader that the anomalous proximity effect refers to the penetration of zero-energy states from the superconductor into the disordered normal metal, which induces a zero-energy peak in the LDOS, driven by the penetration of odd-frequency spin-triplet s-wave pairing. We attribute 
the aforementioned similarity between the trivial and topological phases for long $L_{DN}$  
to the fact that disorder increases the spatial overlap of MBSs, leading to a finite energy splitting. The energy scale of the splitting of LDOS depends on the disorder realization. After ensemble-averaging,  the resulting LDOS 
becomes almost constant within the energy range inside the fan-shaped structure. 

If we focus on the LDOS  as a function of energy $E$ for fixed disorder strength $W$, we see that the ZEP in the trivial phase is very sensitive to disorder. In the clean limit $W=0$, the ZEP is very sharp and has a height similar to that in the topological phase, see the grey curve in Fig.\,\ref{fig4}(d,i). As disorder increases, the ZEP in the trivial phase turns into almost a flat plateau as a function of energy. This is true for all lengths of the DN region. On the other hand, in the topological phase, the ZEP remains robust against disorder for short DN regions albeit with a small height, see the orange and pink curves in Fig.\,\ref{fig4}(i). For long DN regions, the ZEP in the topological phase also turns into a flat plateau as a function of energy, similar to the trivial phase.

If we now look at the spatial profile of the LDOS at $E\approx 0$ for different disorder strengths $W$ and for $L_{\rm DN}=20a$, we see that LDOS in entire system is significantly affected by disorder in both trivial and topological phases.
By comparing the grey and colored curves in Fig.\,\ref{fig4}(e,j), 
we see that disorder significantly reduces the value of the LDOS  
even in the S region despite being completely free of disorder. 
Interestingly, when $W=2\mu_{\rm S}$, the LDOS in the trivial phase slowly decays in the DN region and becomes oscillatory in the clean region, while the LDOS in the topological phase stays flat in the DN region and develops oscillations in the clean region. 

To further understand the nature of the ZEP and the features observed in the LDOS for the short superconducting segment, we now turn our attention to the  the proximity-induced spin-triplet and spin-singlet pair correlations. In Fig.\,\ref{fig5}, we show the absolute value of the spin-triplet pair correlations $|\vec{d}|$ at the leftmost site of the N region ($i=1$) as a function of energy $E$ and disorder strength $W$ for different lengths of the DN region ($L_{\rm DN}$). In the trivial phase (top row of Fig.\,\ref{fig5}), we observe a ZEP in $|\Vec{d}|$ in the clean limit. As disorder is introduced, the peak splits rapidly, evolving into a fan-shaped structure similar to that observed in the LDOS. Taking line cuts at different values of $W$, we clearly see the flattening of the ZEP into almost a plateau as a function of energy. \editE{It is important to clarify why these odd-frequency s-wave triplet pairing, normally robust to scalar disorder, are fragile here. Their fragility is inherited from their parent state; since the trivial ABS is not topologically protected and is easily gapped out by disorder (as we discussed in Sec.\ref{sec4 physical picture}), the pair correlations it induces are consequently suppressed at zero energy. Instead, we have a dip at zero energy with two peaks at finite energy corresponding to the energy aquired by the ABS. Note that, however, due to disorder-averaging, the two peaks are smeared out and become the plateau structure observed here.} The spatial profile of the spin-triplet pairing at $E\approx0$ (Fig. \ref{fig5}(e)\editE{)} shows that in the clean limit (grey curve), triplet correlations exhibit a large value at the DN/S interface where the ABS is localized and it penetrates significantly into the N region. However, with increasing disorder (red and cyan curves), the magnitude of these zero-energy triplet correlations is substantially reduced throughout the entire N region, and their characteristic oscillatory pattern is dampened, especially within the DN segment. 

In the topological phase (bottom row of Fig.\,\ref{fig5}), a ZEP is also present in the clean limit. Moreover, its response to disorder is notably similar to that of the trivial phase. Particularly for a long DN region (Fig.\,\ref{fig5}(g,h)), the splitting and the resulting fan structure become quite pronounced, closely resembling the behavior observed in the trivial phase. This fragility is further confirmed in the line cuts in Fig.\,\ref{fig5}(i). The spatial profile of triplet pairing in Fig.\,\ref{fig5} further illustrates that the induced zero-energy triplet pairing is suppressed by disorder. This suppression is seen in the entire junction including the superconducting region. Overall, the behavior of the spin-triplet pairing in both the trivial and topological phases under disorder matches exactly to that of the LDOS.

To further understand the nature of the pairing, we examine the spin-singlet pairing in Fig.\,\ref{fig6}. A common feature across both the trivial phase (top row) and topological phase (bottom row) is the suppression of spin-singlet pairing around zero energy. Instead of a peak, we observe a dip-like structure as a function of energy. This dip-like structure persists even with increasing disorder. Focusing on the spatial profiles of the spin-singlet pairing evaluated at $E\approx0$ in Fig.\,\ref{fig6}, we see that the spin-singlet component is completely absent in the normal region (both CN and DN segments) for all disorder strengths. Thus, spin-singlet pairing is entirely confined within the S region. This consistent suppression of spin-singlet pairing at zero energy within the N region, irrespective of the phase or disorder strength, stands in stark contrast to the behavior of the spin-triplet components. This observation robustly confirms that conventional even-frequency spin-singlet pairing is not responsible for the ZEP in the LDOS. Instead, the low-energy proximity effect in the N region under these conditions is clearly dominated by unconventional odd-frequency spin-triplet correlations

\subsection{Anomalous Proximity Effect for Long S Regions}
\label{sec4b}
Having established that the ZEP in the LDOS for short $L_{\rm S}$ junctions is mainly due to odd-frequency spin-triplet pairing, and noting its fragility under disorder in both the trivial and topological phases, we now investigate the impact of increasing the length of the S region to $L_{S}=100a$ on the LDOS and pair correlations. This allows us to assess the role of enhanced topological protection due to reduced spatial overlap of MBSs and minimized hybridization energy.

Fig.\,\ref{fig7} displays the LDOS at the N region edge for the $L_{\rm S}=100a$ case. In the trivial phase (top row), the LDOS behavior remains consistent with our findings for the short $L_{\rm S}$ junction (Fig.\,\ref{fig4}). The ZEP observed in the clean limit displays the same fragility against disorder, rapidly splitting and broadening into the fan-shaped structure as $W$ increases, see Fig.\,\ref{fig7}(a,b,c). This is clearly seen in the line cuts at $W=1\mu_S$ and $W=2\mu_S$ in Fig.\,\ref{fig7}(d). The spatial profile at $E\approx0$ (Fig.\,\ref{fig7}(e)) further shows that disorder strongly suppresses the LDOS throughout the N region, indicating that trivial zero-energy ABSs do not acquire enhanced robustness from a longer S segment.

In contrast, the topological phase (bottom row of Fig.\,\ref{fig7}) for long $L_{\rm S}$ junction exhibits a dramatically different response to disorder. The ZEP in the LDOS remains remarkably robust, pinned at $E=0$, even in the presence of moderate to strong scalar disorder, see Fig.\,\ref{fig7}(f,g,h). This is clearly seen in the line cuts in Fig.\,\ref{fig7}(i) where the sharp ZEP persists for $W/\mu_S=1,2$, a behavior not seen in the trivial phase or in the topological phase with a short $L_{\rm S}$. This robustness is a key signature of topologically protected MBSs, whose hybridization is suppressed by the extended length of the S region. Interestingly, when we look at the spatial profile of the LDOS at $E\approx 0$ in Fig.\,\ref{fig7}(j), we see that the LDOS is enhanced at the DN/S interface, suggesting that disorder penalizes MBSs leakage into the N region. We also observe that the LDOS decays exponentially in the DN region appreciably faster than in the short S region case. We verified that this behavior is consistent with the behavior of the wavefunction profile of the MBSs, which becomes more localized at the DN/S interface and decays rapidly into the N region as disorder increases. 

Complementing our results regarding the LDOS in long $L_{\rm S}$ junctions, we present the absolute value of the corresponding induced spin-triplet pair correlations $|\Vec{d}|$ in Fig.\,\ref{fig8}. The behavior of spin-triplet pair correlations further cements the distinction between the trivial and topological phases when MBSs are well-separated. For the trivial phase (top row of Fig.\,\ref{fig8}), the spin-triplet ZEP shows no improved robustness due to longer $L_{\rm S}$, in agreement with what we observed in the LDOS. Conversely, in the topological phase, the spin-triplet pairing ZEP is extremely robust against disorder, similarly to the ZEP in the LDOS. This clearly indicates that the topologically protected MBSs continue to induce substantial odd-frequency spin-triplet pairs that penetrate into the N-region edge. Therefore, the robustness of the ZEP in the LDOS and spin-triplet pair correlations against disorder for long S regions provides a clear distinction between the trivial and topological phases. This suggests that anomalous proximity effect measurements can be used to distinguish between trivial ABSs and topological MBSs, provided that the S region is sufficiently long.

Finally, we now look at the spin-singlet pairing in Fig.\,\ref{fig9}. We see that the spin-singlet pairing is completely suppressed in the normal region for both the trivial and topological phases and for all values of disorder, similar to the short S region case. However, we see that unlike the case of the short S region, the dip-like structure in the spin-singlet pairing at zero energy is only present in the trivial phase. In the topological phase, the spin-singlet pairing is more uniform, showing a parabolic U-shaped structure as a function of energy, rather than a sharp dip. By examining the spatial profile of the spin-singlet pairing at $E\approx 0$ in Fig.\,\ref{fig9}(e,j), we see that the spin-singlet pairing in the trivial phase is significantly larger than that in the topological phase. This is particularly evident at the DN/S interface, where the spin-singlet pairing is enhanced in the trivial phase compared to that in the topological phase. Nevertheless, the continued overall suppression of spin-singlet pairing at $E\approx0$ penetrating into the N region for both phases, even with a long $L_{\rm S}$, reinforces the conclusion that the anomalous ZEP is not of spin-singlet origin.

\section{Conclusion}\label{sec5}
In conclusion, we have investigated the anomalous proximity effect in disordered junctions in the presence of trivial zero-energy Andreev bound states and topological Majorana bound states. In particular, we have considered a clean normal metal/disordered normal metal/superconductor junction based on Rashba semiconductor nanowire model, and studied the effect of scalar disorder and varying lengths of the superconducting and disordered regions on the low-energy spectrum, local density of states, and proximity-induced pair correlations. We have shown that the junction can be tuned to the trivial helical phase with Andreev bound states or the topological phase with Majorana bound states by varying the Zeeman field. We have also shown that  for short superconducting regions, the low-energy spectrum in the trivial and topological phases displays similar behavior in the presence of disorder, making it difficult to distinguish between them. For long superconducting regions, however, the low-energy spectrum in the topological phase is robust against disorder, while that in the trivial phase is very sensitive to disorder.

The central finding of our work lies in the behavior of the zero-energy peak observed in the local density of states. Our investigation of the local density of states and proximity-induced pair correlations in the junction reveal that a zero-energy peak appears in the local density of states in the clean limit for both the trivial helical and topological phases and is closely mirrored by a peak in the magnitude of the induced odd-frequency spin-triplet pair correlations, whereas the even-frequency spin-singlet pair correlations consistently exhibit a dip at zero energy. The crucial distinction between the phases lies in the stability of the zero-energy peak against scalar disorder in the disordered region. In the trivial helical phase, the zero-energy peaks in both the local density of states and triplet pair correlations are fragile and rapidly split with increasing disorder. In the topological phase, the stability of the zero-energy peak depends significantly on the length of the superconducting segment. For short superconducting segments, where Majorana bound states hybridize each other, the zero-energy peak also splits under disorder, making it difficult to distinguish it from the trivial case based solely on these local probes. However, for longer superconducting segments, which is sufficient to suppress Majorana hybridization and enhance topological protection, the zero energy peak remains robustly pinned at zero energy against disorder. This robust zero-energy peak is a direct signature of the anomalous proximity effect, sustained by odd-frequency spin-triplet pair correlations.

Therefore, our findings indicate that while the presence of a zero-energy peak is not exclusive to the topological phase, its robustness against disorder serves as a key indicator. Reliably identifying Majorana bound states via anomalous proximity effect signatures requires systems with sufficiently long superconducting segments to leverage topological protection. Analyzing the stability of the zero-energy peak under varying disorder provides a potential method for distinguishing true Majorana signatures from trivial zero-energy states in clean normal metal/disordered normal metal/superconductor junctions. The consistent suppression of spin-singlet pairing at zero energy further reinforces the unconventional nature of the pairing associated with the zero-energy peak.

\begin{acknowledgements}
We thank S. Ikegaya and N. Shannon for fruitful discussions. E. A. acknowledges financial support from Nagoya University and Mitsubishi Foundation.  Y. T. acknowledges support from JSPS with Grants-in-Aid for Scientific Research  (KAKENHI Grant Nos. 23K17668,24K00583,24K00578,24K00556,25H00609, and 25H00613). 
 J. C. acknowledges financial support from the Carl Trygger’s Foundation (Grant No. 22: 2093), the Sweden-Japan Foundation (Grant No. BA24-0003), and the Swedish Research Council (Vetenskapsr{\aa}det Grant No. 2021-04121). 
\end{acknowledgements}


\printbibliography

\end{document}